\documentclass[reprint, onecolumn,
notitlepage,longbibliography]{revtex4-1}

\usepackage{graphicx} 
\graphicspath{{./figures/}}
\usepackage{float}
\usepackage{array} 
\usepackage{paralist} 
\usepackage{subfig}
\usepackage{fancyhdr} % This should be set AFTER setting up the page geometry
\usepackage{threeparttablex}
\usepackage{amsmath}
\usepackage{mathtools}
\usepackage{xcolor}

\newcommand{\ket}[1]{\ensuremath{\left|#1\right\rangle}} % Dirac Kets
\newcommand{\bra}[1]{\ensuremath{\left\langle#1\right|}} % Dirac Kets
 
\newcommand{\wmc}[1]{\ensuremath{\widetilde{\mathcal{#1}}}}
\renewcommand{\bf}[1]{\ensuremath{\mathbf{#1}}}
\begin{document}
\title{Direct Application of the Phase Estimation Algorithm to Find the Eigenvalues of the Hamiltonians
} 
 \author{Ammar~Daskin}
 %\email{also known as Anmer Daskin. email: adaskin25@gmail.com}
\affiliation{Department of Computer Engineering, Istanbul Medeniyet University, Kadikoy, Istanbul, Turkey}
 \author{Sabre~Kais}
\affiliation{Department of Chemistry, Department of Physics and Birck Nanotechnology Center, Purdue University, West Lafayette, IN, USA}
\affiliation{Qatar Environment and Energy Research Institute, HBKU, Doha, Qatar} 
\begin{abstract}
The eigenvalue of a Hamiltonian, $\mathcal{H}$, can be estimated through the phase estimation algorithm given the matrix exponential of the Hamiltonian, $exp(-i\mathcal{H})$. 
The difficulty of this exponentiation impedes the applications of the phase estimation algorithm particularly when $\mathcal{H}$ is composed of non-commuting terms.
In this paper, we present a method to use the Hamiltonian matrix directly in the phase estimation algorithm by using an ancilla based framework: 
In this framework, we also show how to find the power of the Hamiltonian matrix-which is necessary in the phase estimation algorithm-through the successive applications. 
This may eliminate the necessity of matrix exponential for the phase estimation algorithm and therefore  provide an efficient way to estimate the eigenvalues of particular Hamiltonians.
The classical and quantum algorithmic complexities of the framework are analyzed for the Hamiltonians which can be written as a sum of simple unitary matrices and shown that a Hamiltonian of order $2^n$ written as a sum of $L$ number of simple terms can be used in the phase estimation algorithm with $(n+1+logL)$ number of qubits and 
$O(2^anL)$ number of quantum operations, where $a$ is the number of iterations in the phase estimation. 
In addition, we use the Hamiltonian of the hydrogen molecule as an example system and present the simulation results for finding its ground state energy.
\end{abstract}

\maketitle

\section{Introduction}
 With the recent efforts such as digitalizing adiabatic quantum computers \cite{Barends2016digitized}, a digital quantum simulation of the real-time dynamic of a lattice gauge theory\cite{Martinez2016real}, a simulation of the  Hubbard model\cite{Salfi2016quantum}, and the race to build universal quantum computers among the big companies \cite{Castelvecchi2017quantum}, 
 quantum computers become closer to be used in unsolved real-world applications.
 
As Feynman suggested\cite{Feynman1982simulating}, one of the breakthroughs of  quantum computers is expected to be in the simulation of  quantum systems, a known difficult problem for classical computers because of the exponential growth of the system size with the number of qubits 
(e.g., see Ref.\cite{Troyer2005computational}). 
 In the standard formalism of quantum mechanics, the Hamiltonian of a quantum system is considered as a Hermitian matrix.  
 Since the exponential of a skew-Hermitian matrix is  a unitary matrix, the time evolution, $exp(-i\hbar\mathcal{H})$, of a quantum system represented by the Hamiltonian $\mathcal{H}$ describes a unitary transformation.
 In general, a computation or an algorithm should be represented by a unitary matrix in order to be implemented on a quantum computer.
 If an algorithm can be formalized in terms of unitary matrices, then a possible implementation may yield a computational efficiency over the classical algorithms: e.g. Shor's integer factoring algorithm \cite{Shor1994algorithms}.
Many quantum algorithms use the exponential  $exp(-i\hbar\mathcal{H})$ to map the description of the original problem 
$\mathcal{H}$ to a unitary matrix.
An exact mapping through matrix exponential generally requires eigendecomposition of $\mathcal{H}$: In spite of the existence of various numerical methods, the eigendecomposition in the cases where 
$\mathcal{H}$  is a large-dense matrix is still a big computational challenge for classical computers.
Because of this computational difficulty, it is common to use some order of Trotter-Suzuki approximation\cite{Suzuki1976generalized} by writing $\mathcal{H}$ as a sum of terms whose exponentials are known or easy to compute. 
In general case,  the order of the approximation affects the correctness and the high order approximations dramatically increases the computational complexity \cite{Poulin2015TSS}. 
Therefore,
the complexity of this exponentiation should be also embodied in the consideration of the computational complexity of a quantum algorithm (see Ref.\cite{Scott2015Caveats} for a relevant discussion of the topic). 

It is known that using additional qubits eases  the implementation difficulty of a quantum circuit. 
A quantum operator (or a dynamic of a system) can be implemented inside a larger system, where some reduced part of the system represents the action of the original system.
This idea  is used in different contexts such as designing circuits from matrix elements \cite{Daskin2012universal,Daskin2014universal}, 
and generating exponential of a matrix written as a sum of unitary matrices through Taylor series \citep{Berry2015Taylor}. Moreover, in Ref.\cite{Daskin2017ancilla}, to use a matrix on a quantum computer directly,  we considered converting a non-unitary matrix, 
$\mathcal{H}$, to a unitary by using its square root.
In a more general fashion, Ref.\cite{Low2016hamiltonian} described a qubitization approach to complete a non-unitary matrix to a unitary matrix whose multiple applications generate Chebyshev polynomials of $\mathcal{H}$.

In this paper, we follow a different approach: for a Hamiltonian $\mathcal{H}$ of order $2^n$ with real
 eigenvalues; 
  we first put the matrix in the form $\wmc{H}=\left(\bf{I} - i\mathcal{H}/\kappa\right)$,  where $\bf I$ is an identity matrix and $\kappa$ is a coefficient. 
For  a sufficiently large $\kappa$; 
the value of the angle, the phase, seen in the polar form of the eigenvalue of \wmc{H} becomes approximately  equal to $\lambda_j/\kappa$, where $\lambda_j$ is an eigenvalue of $\mathcal{H}$.
Using this fact and assuming $\mathcal{H}$ is a sum of some simple terms which can be easily mapped to quantum circuits; 
 we present a phase estimation framework for finding $\lambda_j/\kappa$. 
 In this framework, we use \wmc{H} directly and show that necessary powers of $\wmc{H}$ can be estimated through successive applications after the application of the oblivious amplitude amplification. 
This provides a way to estimate the  eigenvalues of $\mathcal{H}$  by using $(n+1+logL)$ qubits and $O(2^anL)$ number of quantum operations.
 Here, $a$ is the number of iterations in the phase estimation and $L$ is the number of terms in the sum yielding $\mathcal{H}$.
We also show that the powers of $\wmc{H}$ can be more accurately estimated by using  additional qubits and a permutation operator in the phase estimation algorithm, which requires also the same number of quantum gates and $O(a+n+logL)$  number of qubits.   
As an example system, we use the Hamiltonian matrix of the hydrogen molecule and estimated its ground state energy within the presented framework. 

In the following sections, after first describing the proposed framework, we will present the complexity analysis. 
Then, in Sec.\ref{SecH2}, the simulation results for the hydrogen molecule are presented and discussed. And in the final section the paper is concluded with a summary.

\section{Proposed Eigenvalue Estimation}
 Phase estimation algorithm (PEA) estimates the phase $\phi$ for a given operator with the eigenvalue $e^{i\phi}$. 
 For a Hamiltonian $\mathcal{H} \in C^{\otimes n}$, we propose to consider the following matrix in the phase estimation algorithm:
 \begin{equation}
 \label{EqBaseForm}
\wmc{H} = \left(\bf{I} - \frac{i\mathcal{H}}{\kappa}\right)
 \end{equation}
where $\bf I$ is an identity matrix, and $\kappa$ is a coefficient.
 The eigenvalues of this matrix are in the form 
 $\widetilde{\lambda}_j = ( 1-i\lambda_j/\kappa)$, where $\lambda_j$ is the $j$th eigenvalue of $\mathcal{H}$.   When $\kappa$ is sufficiently large; in the polar form of  $\widetilde{\lambda}_j$,  
 the value of the angle  becomes $\lambda_j/\kappa$ since $sin(\lambda_j/\kappa)\approx \lambda_j/\kappa$:
\begin{equation}
 \widetilde{\lambda}_j= 1-i\frac{\lambda_j}{\kappa} = \left|1+i\frac{\lambda_j}{\kappa}\right|e^{i\frac{\lambda_j}{\kappa}}.
\end{equation}
Also note that when $\kappa$ is large, $\wmc{H}^\kappa$ gives an approximation to $e^{i\mathcal{H}}$.
To be able to use the matrix in Eq.\eqref{EqBaseForm} directly in the phase estimation, we will initially assume that we have an efficient mechanism to produce the following unitary matrix:
\begin{equation}
\label{EqU(1)}
\bf{U}^{(1)}
= \left(\begin{matrix}
\wmc{H}  & \bullet\\
\bullet&\bullet
\end{matrix}\right),
\end{equation}
where the first part of the matrix is equal to \wmc{H} with/without some normalization, and a $\bullet$ indicates another part of the matrix.
As also mentioned in the introduction, these types of matrices are used in various contexts: either to design quantum circuits from matrix elements \cite{Daskin2012universal,Daskin2014universal} or to be able to use it with the oblivious amplitude amplification\cite{Daskin2017ancilla}, or to estimate unitary dynamic of a Hamiltonian through truncated Taylor series \cite{Berry2015Taylor} and its generalized form \cite{Low2016hamiltonian}. 
In the next section we will discuss how the above matrix can be generated by using an ancilla register. 
Now let us consider the input state \ket{\bf0}\ket{\varphi}, where \ket{\varphi} represents an abstract quantum state on the system register and  \ket{\bf0} is the first vector in the standard basis and the state on the ancilla register. The application of $\bf{U}^{(1)}$ to this input yields the following state \cite{Berry2015Taylor}:
\begin{equation}
\label{EqOutputU}
\bf{U}^{(1)}\ket{\bf{0}}\ket{\varphi} = 
{\sqrt{p_0}}\ket{\bf{0}} \wmc{H}\ket{\varphi} 
+ 
\sqrt{1 - p_0}\ket{\varphi}.
\end{equation}
The application of \wmc{H} to the input \ket{\varphi} on the system register can be obtained  from the above output with the probability given by $p_0||\wmc{H}\ket{\varphi}||$, where $||.||$ represents a vector norm. 
For a unitary \wmc{H}, the probability simply becomes  $p_0$.
The desired output can be singled out through the application of a projector $\bf{ P} = \ket{\bf{0}}\bra{\bf{0}}$ to the ancilla register.
 \subsection{Estimation of $\wmc{H}^{2^j}$}

 For the phase estimation, we  need to be able to efficiently generate the set of operators $\{\bf{U}^{(2^0)}\dots \bf{U}^{(2^a)}\}$ which consist of $\wmc{H}^{2^0} \dots \wmc{H}^{2^a}$. 
 This  can be done through the successive applications of $\bf{U}^{(1)}$ after the application of the oblivious amplitude amplification described in Sec.\ref{SecOblivousAA}: The amplitude amplification process to maximize the probability of \ket{\bf 0} in the ancilla  turns $\bf{U^{(1)}}$ into the following form:
 \begin{equation}
\label{EqU(1)AA}
\bf{\widetilde{U}}^{(1)}
\approx \left(\begin{matrix}
\wmc{H}  &  0\\
 0&\bullet
\end{matrix}\right),
\end{equation} 
$k$ number of applications of the matrix $\bf{\widetilde{U}}^{(1)}$ can be used to obtain  $k$th power of $\wmc{H}$: 
\begin{equation}
\underbrace{\bf{\widetilde{U}}^{(1)}\dots\bf{\widetilde{U}}^{(1)}}_{k \text{ times}} 
\approx \left(\begin{matrix}
\wmc{H}^k  &  0\\
 0&\bullet
\end{matrix}\right).
\end{equation}

A more accurate estimation for the $k$th power of   $\wmc{H}$ in $\bf U^{(k)}$ can be  obtained from $\bf U^{(k/2)}$ by using an additional qubit in the ancilla and a permutation matrix: 
 For instance, if $\wmc{H}^2$ is desired, we first add one more qubit to the ancilla. Then the matrix representation of the circuit becomes
\begin{equation}
\bf{I^{\otimes1}}\otimes\bf{U}^{(1)}=\left(\begin{matrix}
\wmc{H}  & \bullet & \bf{0} & \bf{0}\\
\bullet&\bullet& \bf{0} & \bf{0}\\
\bf{0} & \bf{0}&\wmc{H}& \bullet  \\
\bf{0} & \bf{0}&\bullet&\bullet
\end{matrix}\right).
\end{equation}
Using a permutation matrix, $\bf{\Pi}$, similar to the one below, we can obtain the square from a successive application of this matrix as shown below:
\begin{equation}
\label{EqPowerofH}
\left(\bf{I^{\otimes1}}\otimes\bf{U}^{(1)}\right) 
\left(\begin{matrix}
\bf{I}  & \bf{0} & \bf{0} & \bf0\\
\bf0&\bf0& \bf{I} & \bf0\\
\bf{0} & \bf{I}&\bf{0}& \bf{0}  \\
\bf{0} & \bf0&\bf{0}&\bf{I}
\end{matrix}\right)
\left(\bf{I^{\otimes1}}\otimes\bf{U}^{(1)}\right)
=
\left(\begin{matrix}
\wmc{H}^2  & \bullet & \bullet &\bullet\\
\bullet&\bullet& \bullet & \bullet\\
\bullet&\bullet& \bullet & \bullet\\
\bullet&\bullet& \bullet & \bullet
\end{matrix}\right),
\end{equation}
where the dimension of the sub-matrices are assumed to conform to the multiplications. 
In more general sense, 
$\bf{U^{(k)}} 
= (\bf{I}^{\otimes log k}\otimes 
\bf{U^{(k/2)}})\bf{\Pi} (\bf{I}^{\otimes log k}\otimes \bf{U^{(k/2)}})$. Here,  $\bf{\Pi}$ can be implemented through some swap operations on the ancilla register.

\begin{figure}
\includegraphics[scale=1]{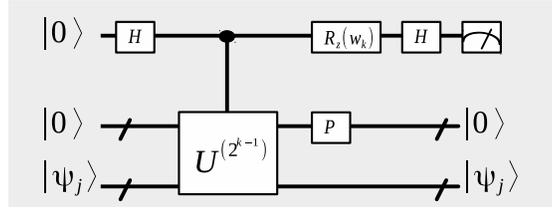}
\caption{The $k$th iteration of the phase estimation algorithm: In the circuit, \ket{\psi_j} is an approximate eigenvector of $\mathcal{H}$ and
$w_k = -2\pi(0.0\phi_k\phi_{k-1}\dots \phi_a)$, where $\phi_k, \phi_{k-1},\dots ,\phi_a$ represents the previously measured bit values.
\label{FigIPEA}
}
\end{figure}

Consequently,  we can use the set of operators $\{\bf{U}^{(2^0)}\dots \bf{U}^{(2^a)}\}$ generated by successive applications in the phase estimation algorithm and estimate the value of $\frac{\lambda_j}{\kappa}$  and hence the eigenvalue of the Hamiltonian with an accuracy affected by $a$. 
Because of the additional qubits, it may be easier to implement this method through the iterative version of the phase estimation algorithm as depicted in 
Fig.\ref{FigIPEA}.
\subsection{The steps inside PEA}
The phase estimation algorithm at the $x=(a-k-1)$th iteration is composed of three registers: vis., the phase with one-qubit, the ancilla with $x+logL$ qubits, and the system register with $n$ qubits.
The initial input to the algorithm is given by the following:
\begin{equation}
\ket{\bf\psi_0} = \ket{0}\ket{\bf0}\ket{\bf\varphi_j},
\end{equation}
where \ket{\bf\varphi_j} is an estimation to the $j$th eigenvector of 
$\mathcal{H}$. 
In PEA, we first apply the Hadamard gate to the phase qubit which simply puts this qubit into the superposition and yields  the state:
\begin{equation}
\ket{\bf\psi_1} = \frac{1}{\sqrt{2}}
\left(\ket{0}\ket{\bf0}\ket{\bf\varphi_j} + \ket{1}\ket{\bf0}\ket{\bf\varphi_j}\right).
\end{equation} 
Then, $\bf{U^{(2^k)}}$ controlled by the phase qubit is applied to the remaining qubits:
\begin{equation}
\ket{\bf\psi_2} = \frac{1}{\sqrt{2}}
\left(\ket{0}\ket{\bf0}\ket{\bf\varphi_j} + \ket{1}\bf{U^{(2^k)}}\ket{\bf0}\ket{\bf\varphi_j}\right).
\end{equation} 
With the help of Eq.\eqref{EqOutputU}, \ket{\bf\psi_2} can be rewritten in the following form:
\begin{equation}
\begin{split}
\ket{\bf\psi_2} = & \frac{1}{\sqrt{2}}
\left(\ket{0}\ket{\bf0}\ket{\bf\varphi_j}\right)
 \\ & 
 + 
 \frac{\ket{1}}{\sqrt{2}}
\left(
\ket{\bf{0}} \hat{p}
e^{i2\pi \phi_j2^k}\ket{\bf\varphi_j}
+ 
\sqrt{1 - \hat{p}^2}
\ket{\Phi}\right),
\end{split}
\end{equation}
where $\hat{p} = \sqrt{p_0}|\widetilde{\lambda}_j|^{2^k}$.
The application of the operator $\bf{P}$ to the ancilla register causes the state to collapse into the following \underline{unnormalized} state:
\begin{equation}
\ket{\bf\psi_3} =  \frac{1}{\sqrt{2}}
\left(\ket{0}\ket{\bf0}\ket{\bf\varphi_j}
 +
\hat{p}\ket{1}\ket{\bf{0}} 
e^{i2\pi \phi_j2^k}\ket{\bf\varphi_j}
\right).
\end{equation}
In the iterative phase estimation, in each iteration we get the estimate $k$th bit value of the phase. 
This is achieved with the help of a $R_z$ gate whose rotation angle is determined from the previous iterations (please refer to Fig.\ref{FigIPEA} for the value of the angle.). 
This gate  converts the term $e^{i2\pi \phi 2^k}$ into the form $e^{i2\pi (0.x_k)}$, where $x_k$ represents a binary bit: i.e., simply:
\begin{equation}
\ket{\bf\psi_3} =  \frac{1}{\sqrt{2}}
\left(\ket{0} + 
\hat{p}\ket{1}
e^{i2\pi(0.x_k)}
\right)\ket{\bf0}\ket{\bf\varphi_j},
\end{equation} 
In the final step, the Hadamard gate is applied again to the phase qubit:

\begin{equation}
\begin{split}
\ket{\bf\psi_{final}} = &  \frac{1}{2}
\left(
1+\hat{p} e^{i2\pi(0.x_k)}\right)
\ket{0} \ket{\bf0}\ket{\bf\varphi_j}
\\ & + 
\frac{1}{2}\left(
1-\hat{p} e^{i2\pi(0.x_k)}
\right)
\ket{1} \ket{\bf0}\ket{\bf\varphi_j}
\end{split}
\end{equation} 
Measurement on the phase qubit in this unnormalized state yields the bit value of $x_k$:
\begin{itemize}
\item When $x_k = 0$, the probability of \ket{0} becomes greater in the final state:
\begin{equation}
  \frac{1}{2}
\left(\left(
1+\hat{p}\right)\ket{0}
 + \left(
1-\hat{p} 
\right)\ket{1}\right)
 \ket{\bf0}\ket{\bf\varphi_j}
\end{equation} 
\item In the case $x_k = 1$, the probability of \ket{1} becomes greater:
\begin{equation}
  \frac{1}{2}
\left(\left(
1-\hat{p}\right)\ket{0}
 + \left(
1+\hat{p} 
\right)\ket{1}\right)
 \ket{\bf0}\ket{\bf\varphi_j}
\end{equation} 
\end{itemize} 
Here, the probability difference between measuring 1 and 0 is determined by 
$\frac{2\hat{p}}{\mu }$, where $\mu$ is a normalization constant.
Before the application of the projector $\bf{P}$,  $\mu$  can also be eliminated by applying the  oblivious amplitude amplification to $\bf{U}^{(1)}$ as  described below. 

\subsection{Application of oblivious amplitude amplification}
\label{SecOblivousAA}
The amplitude amplification \cite{Mosca1998quantum, Brassard2002,Kaye2006} is based on the Grover's search algorithm  \cite{Grover1998} where one applies a sequence of the operators to increase the magnitude of the amplitudes of some desired states: Consider the following output state:
 \begin{equation}
\bf{A}\ket{\bf0} = \sum_{x\in X_{good}}\alpha_x\ket{\bf x}\ket{\bf\Phi} + \sum_{x\in X_{bad}}\alpha_x\ket{\bf x}\ket{\bf\Phi},
\end{equation}
where $\bf{A}$ is a quantum algorithm,  $X_{good}$ and  $X_{bad}$ are the sets of good (desired) states and bad(undesired) states of the first register, $x$ represents a standard basis vector, and \ket{\bf\Phi} represents the  states of the qubits in the second register. 
The probability of the good states in this output can be increased by the application of the iteration operator: 
\begin{equation}
\label{EqAAmp}
\bf{Q} = \bf{A}\bf{U_0}^\perp \bf{A}^\dagger \bf{U_f}.
\end{equation}
Here, the operator $\bf{U_f}$  marks (multiply by -1) the amplitudes of the desired states and 
$\bf{U_0}^\perp = 2\ket{\bf 0}\bra{\bf 0} - \bf{I}$. 

A version of the amplitude amplification called oblivious amplitude amplification \cite{Paetznick2014repeat,Berry2014exponential,Berry2015Taylor}  can be used to increase the probability of the part $\left({\sqrt{p_0}}\ket{\bf{0}} \wmc{H}\ket{\varphi}\right)$ in Eq.\eqref{EqU(1)}. 
In the oblivious amplitude amplification  algorithm, $\bf{U_f}$ and $\bf{U_0}^\perp$ only operates on the ancilla 
and the probability change does not affect the state in the second register. In our case, we use the following iteration operator: 
\begin{equation}
\label{EqOAAmp}
\bf{Q} = \bf{U}^{(1)}
\left(\bf{U_0}^\perp \otimes \bf{I}^{\otimes n}\right) 
{\bf{U}^{(1)}}^\dagger 
\left(\bf{U_0}^\perp \otimes \bf{I}^{\otimes n}\right),
\end{equation}
where $\bf{U_0}^\perp$ acts on the ancilla register. 
The main difference from  Eq.\eqref{EqAAmp} 
 is that Eq.\eqref{EqOAAmp} does not depend on the input to the system register and the marking and the amplifying  operators, $\bf{U_0}^\perp,$ are applied only on the ancilla register.

The oblivious amplitude amplification algorithm best works  when $\wmc{H}$ is a unitary process. However, 
recently it is shown that the algorithm also works when $\wmc{H}$ is close to a unitary matrix \cite{Berry2015Taylor}.
In this case, the error in the output of one step of the oblivious amplification is bounded by the distance which in our case measured as: 
 $\|\wmc{H}\wmc{H}^\dagger - \bf{I}\|$.
Since $\wmc{H}\wmc{H}^\dagger = \bf{I} + \mathcal{H}^2/\kappa^2$, this distance becomes:
\begin{equation}
  \|\wmc{H}\wmc{H}^\dagger - \bf{I}\| = \left\|\wmc{H}^2/\kappa^2\right\|.
\end{equation}
 Thus, the error in the output of the oblivious amplitude amplification is bounded by $O(\|\wmc{H}^2/\kappa^2\| )$.
 For a large $\kappa$, $\| \mathcal{H}^2/\kappa^2\|$ and so the order of the error can be expected to be very small.
In addition,  since $\wmc{H}^\kappa$ gives an approximation to $e^{i\mathcal{H}}$, in the iterations of PEA,  $\wmc{H}^{2^j}$ for some $j$ can be expected to be nearly a unitary matrix.
Therefore, in the case of \wmc{H}  with a large $\kappa$, 
it is possible to use oblivious amplitude amplification.
To observe this,  the simulation results in Sec.\ref{SecH2} presented with and without the oblivious amplitude amplification.
 %%%%%%%%%%%%%%%%%%%%%%%%%%%%%%%%%%%%%
\subsection{Simulating sums of unitary matrices} 
Any Hamiltonian can be decomposed into a linear combination of unitary matrices:
\begin{equation}
\label{HsumofHl}
\mathcal{H} = \sum_{l=1}^L\alpha_l\bf{H_l},
\end{equation} 
where $\alpha_l$ is some complex coefficient and $\bf{H_l}$ represents a unitary matrix. 

Berry et al.\cite{Berry2015Taylor} have showed 
that the time evolution of $\mathcal{H}$ can be simulated as follows:
The time is considered to be divided into $r$-segments  so that the time evolution at segment $r$, 
$\bf{U_r} = e^{-i\mathcal{H}t/r}$, is approximated through the following Taylor expansion:
\begin{equation}
\bf{U_r} \approx \sum_{k=0}^K\frac{1}{k!}(\frac{-i\mathcal{H}t}{r})^k.
\end{equation}
Since $\mathcal{H}^k=\sum_{l1,\dots, l_k =1}^L\alpha_{l_1}\dots\alpha_{l_k}\bf{H_{l_1}}\dots \bf{H_{l_k}}$, this expansion is rewritten as:
\begin{equation}
\bf{U_r}  \approx 
\sum_{k=0}^K \frac{(-it/r)^k}{k!}
\sum_{l1,\dots, l_k =1}^L\alpha_{l_1}
\bf{H_{l_1}} \dots \alpha_{l_k}\bf{H_{l_k}}
 = \sum_{j=0}^M \beta_j \bf{V_j},
\end{equation} 
where $\bf{V_j}$ is in the form of $(-i)^k\bf{H_{l_1}}\dots \bf{H_{l_k}}$, $\beta_j$ is a complex coefficient, and $M$ is the number of terms in the summation and equal to $L^k$.
Assuming a mechanism to implement each $\bf{V_j}$ exist, 
then $\bf{V_j}$s are combined in a quantum circuit $select(\bf{V})$ by using $m = logM$ ancilla qubits in a way that for each state $\ket{\bf{j}}$ on the ancilla,  a $\bf{V_j}$ is applied to the main qubits.
Here,  $\ket{\bf{j}}$ represents the $j$th vector in the standard basis. 
For an arbitrary state \ket{\psi}, this can be shown as:  
 \begin{equation}
 select(\bf{V})\ket{j}\ket{\psi}=
\ket{j}\bf{V_j}\ket{\psi}.
 \end{equation}
The state on the ancilla can be put into a superposition state where each basis biased by the coefficients $\beta_j$s: 
\begin{equation}
\bf{B} \ket{\bf{0}} = 
\sqrt{p_0}\sum_{j=0}^M\sqrt{\beta_{j}}\ket{j};
\end{equation} 
where $p_0 = \sum_{j=0}^M \beta_j$ and  $\beta_j$s are assumed to be positive.
The time evolution $\bf{U_r}$ can be obtained by combining the operator $select(\bf{V})$ with $\bf{B}$  in the following way: 
\begin{equation}
\label{EqBSelectVB}
\bf{U}=(\bf{B}^\dagger 
\otimes \bf{I})
select(\bf{V})(\bf{B}\otimes \bf{I})= \left(\begin{matrix}
\bf{U_r} & \bullet\\
\bullet&\bullet
\end{matrix}\right), 
\end{equation}
As similarly explained in the previous sections, this matrix can be used simulate action of $\bf{U_r}$ on some arbitrary state \ket{\psi} with the probability determined by $p_0$.

When $\mathcal{H}$ given in the form of Eq.\eqref{HsumofHl}, 
the matrix described in Eq.\eqref{EqBaseForm} can also be represented as a sum of unitary matrices:
\begin{equation}
\wmc{H}=\left(\bf{I} - \frac{i\mathcal{H}}{\kappa}\right) = 
\frac{1}{\kappa}\sum_{l=0}^L -i\alpha_l\bf{H_l} = \sum_{l=0}^L \beta_l \bf{V_l},
\end{equation}
where $\bf{H_0} = i\kappa\bf{I}$.
Using the same methodology given in Eq.\eqref{EqBSelectVB}, we can form the matrix $\bf{U}^{(1)}$ in Eq.\eqref{EqU(1)} easily.

Also note that since $\wmc{H}$ is a sum of unitary matrices, its powers are also similar to those of $\mathcal{H}$. For instance, its $k$th power reads as follows:
\begin{equation}
\wmc{H}^k = \frac{1}{\kappa^k}
\sum_{l1,\dots,l_k = 0}^L (-i)^k\alpha_{l1}\bf{H_{l_1}}\dots \alpha_{lk}\bf{H_{l_k}}.
\end{equation}
Therefore, 
\begin{equation}
\bf{U}^{(k)}
=
 \left(\begin{matrix}
\wmc{H}^k  & \bullet\\
\bullet&\bullet
\end{matrix}\right). 
\end{equation}
However,  in terms of complexity,  the required number of qubits is $klogL$ which grows linearly with the power. 
 In the case of using this matrix inside PEA, the required number of qubits grows exponentially with the precision: 
 That means if we need $\bf{U}^{2^j}$ at an iteration of PEA, we need to use $2^jlogL$ qubits to represent $L^{2^j}$ number of terms in the matrix $select(\bf{V})$. 
 Therefore, when the available number of qubits is limited, the power should be taken through successive applications of Eq.\eqref{EqU(1)AA} or Eq.\eqref{EqPowerofH}. 
 In the case of Eq.\eqref{EqPowerofH}, only one additional qubit is necessary to find the square of  \wmc{H} from the previous iteration of the phase estimation algorithm.
                                                  
\subsection{Simulating sum of rank one matrices}
Any matrix can also be written as a sum of rank one matrices. Consider the matrices in the following form:
\begin{equation}
\label{eqsumofxj}
\mathcal{H} = \sum_j \bf{H_j} = \sum_j \ket{\bf{x_j}}\bra{\bf{x_j}},
\end{equation}
where $\bf{H_j} = \ket{\bf{x_j}}\bra{\bf{x_j}} $ is a rank one Hermitian matrix constructed by the normalized vector $\ket{\bf{x_j}}$.
Such sums  are frequently seen in the form $\bf{XX^T}$ in machine learning problems and statistical analysis of data points-where a column of the matrix $\bf{X}$ represents a data point.
We can rewrite Eq.\eqref{eqsumofxj} as follows:
\begin{equation}
\begin{split}
\mathcal{H} & =  -\frac{1}{2}\sum_{j=1}^L \left[ (\bf{I} - 2\ket{\bf{x_j}}\bra{\bf{x_j}} ) - \bf{I} \right] 
\\ 
& =  -\left(\sum_{j=1}^L \frac{\bf{R_j}}{2}\right) + \frac{L}{2}\bf{I} = \sum_{j=0}^L \alpha_j\bf{R_j},
\end{split}
\end{equation}
where $\alpha_j$s are coefficients and $\bf{R_0} = \bf{I}$.
The above equation describes $\mathcal{H}$ in terms of a sum of $(L+1)$ number of idempotent-unitary matrices:
i.e., $\bf{R_j}^2 = \bf{I}$.
As a result, $\mathcal{H}$ can be simulated in the phase estimation after being turned into the form of Eq.\eqref{EqBaseForm}.
In the next section, we will analyze the complexity of the method in the case the Hamiltonian is given as a sum of unitary matrices.
                                 
\section{Computational Complexity}
Computational complexity of a quantum algorithm involves two aspects: vis., classical and quantum complexities. The classical complexity covers the preprocessing time of circuits implementing the quantum algorithm, i.e. $e^{i\mathcal{H}}$, and post-processing of the output for obtaining a desired solution. 
In this paper, we have mainly focused on avoiding  the necessity of $e^{i\mathcal{H}}$ in estimation of the eigenvalue of $\mathcal{H}$ on quantum computers.  When the matrix $\mathcal{H}$  can be mapped to quantum circuits easily (e.g., it is a sum of "simple" unitary matrices); then the classical complexity is bounded by the number of matrix elements: i.e., $O(N^2)$, assuming all the matrix elements are needed to be processed and stored.  

 On the other hand, quantum complexity  of a quantum algorithm is determined by the number of qubits and the number of one-and two-qubit quantum gates involved in the circuit implementing the quantum algorithm (Here, we only consider algorithmic complexity.).
 In this perspective, the complexity analysis of the method is given below.
\subsection{The number of qubits}
When the powers of \wmc{H} is obtained through Eq.\eqref{EqU(1)AA}, the required number of qubits is $(n+1+logL)$. 
However, when obtained through Eq.\eqref{EqPowerofH},  
it becomes $(a+n+logL)$, where $a$ is the number of iterations in the phase estimation.

\subsection{Circuit Implementation of $\bf{B}$}

For a normalized $L$-dimensional column-vector $\ket{\bf{x_j}} \in C^{\otimes l}$ and the identity matrix $\bf{I}$, $\bf{R_j}=\bf{I}-2\ket{\bf{x_j}}\bra{\bf{x_j}}$ is an Householder transformation describing a reflection operator around the vector $\ket{\bf{x_j}}$. 
On quantum computers, a  Householder transformation \cite{Golub2012matrix} 
 can be implemented by using $O(2^l)$ total number of two- and one-qubit quantum gates \cite{Ivanov2006engineering,Bullock2005asymptotically,Urias2015householder} ( In Ref.\cite{Ivanov2006engineering}, an implementation with the same complexity is also presented for a general version of the Householder transformation: i.e., $\bf{I}-(e^{i\varphi}-1)\ket{\bf{x_j}}\bra{\bf{x_j}}$).

Note that a Householder matrix, $\bf{R}$,  can be also implemented through $(2L-3)$ number of plane (Givens) rotations\cite{Golub2012matrix} as follows: 
It is known that a Givens rotation $\bf{G_{j-1}}$ can be used to zero out 
$j$th  entry in the column of a matrix. 
The Givens rotation $\bf{G_{L-1}}$ applied to $\bf{R}$ not only nullifies the last element in the first column of $\bf{R}$ but all the entries on the $L$th row upto the first sub-diagonal entry as well. 
Furthermore, the matrix product 
$(\bf{G_{L-1} R G_{L-1}^T})$ 
yields a matrix where all entries but the diagonal entry on the last row and column are zeros and the diagonal element is one.
 Because of this property, $\bf{R}$ can be shown as a product of $(2L-3)$ number of plane rotations: That is, $\bf{R} = \bf{G_{L-1}^T}\dots \bf{G{3}^TG_{2}^TG_{1}^T G_{2}G_{3}}\dots \bf{G_{L-1}}$.
 As a result, the circuit for $\bf{R}$ can be formed via plane rotations in this decomposition.   
Since a rotation matrix can be implemented through a multi controlled quantum gate \cite{Cybenko2001reducing};  
when considered together, the rotations in the product  form a uniformly controlled gate-network. 
These networks are well-studied in Ref.\cite{Vartiainen2004efficient} where it is shown that a uniformly controlled network acting on $l$ qubits  can be simplified into $O(2^l)$ number of one- and two-qubit gates. 
Also note that the sparsity of the vector $\ket{\bf{x_j}}$ directly affects the number of plane rotations in the decomposition and hence the number of gates required for the implementation.

\subsection{Circuit Implementation of $select(\bf{V})$}
The circuit for $select(\bf{V})$ is determined by the number of terms which is given by $L$ and the number of gates required to implement each term. 
If each term in the sum can be implemented through simple $O(n)$ quantum gates on different qubits, then the combination of the all terms form $O(n)$ different gray-coded networks controlled by $l$ number of qubits in the ancilla. 
The decomposition of this networks will require $O(nL)$ quantum gates. 

Therefore, the total complexity   to implement 
$\bf{U^{(1)}} = \bf{B}^\dagger 
\otimes \bf{I})
select(\bf{V})(\bf{B}\otimes \bf{I})$  is bounded by $O(nL+2L)=O(nL)$ with the assumption that each term in the sum can be implemented with $O(n)$ quantum gates.
In an iteration of the phase estimation, finding the power $\bf{U^{(2^k)}}$ requires $2^k$ number of successive applications of $\bf{U^{(1)}}$.
Therefore, the total required quantum gates for an iteration of the algorithm  becomes $O(2^knL)$, where $1\leq k\leq a$ with $a$ being the total number of iterations.

\section{An Example Application to the Hamiltonian of $H_2$}
\label{SecH2}
The concept of the second quantization used in quantum chemistry to simplify the formalism of fermionic many particle systems. It represents the interacting systems of electrons and nuclei through the creation and annihilation operators. 
The molecular electronic Hamiltonian in electronic structure problem is expressed in the second quantization form as follows\cite{Lanyon2010towardsH2,Whitfield2011simulation,Seeley2012}:
\begin{equation}
\label{h2hamiltonian}
\mathcal{H}=\sum_{pq}{h_{pq}a_{p}^{\dagger}a_q}+\frac{1}{2}\sum_{pqrs}{h_{pqrs}a_{p}^\dagger a_{q}^\dagger a_{s}a_{r}},
\end{equation}
where $h_{pq}$ is the one-electron integrals including the electronic kinetic energy and the electron nuclear attraction terms. $h_{pqrs}$ represents the set of two-electron integrals with the electron-electron interactions. $a_j$ and $a_j^\dagger$ are the lowering and raising operators. This type of Hamiltonians can be represented in terms Pauli matrices by using the following Jordan-Wigner transforms: 
\begin{equation}
\label{annihilationcreation}
\begin{split}
a_j\rightarrow 
 \sigma_{-}^{j} \left(\prod_{k=1}^{j-1}{\sigma_{z}^{k}}\right), 
 \text{\ and\ } 
a_j^\dagger\rightarrow \sigma_{+}^{j} \left(\prod_{k=1}^{j-1}{\sigma_{z}^{k}}\right),
\end{split}
\end{equation}
where \begin{equation}
 \sigma_{+} = \frac{\sigma_{x}-i\sigma_{y}}{2}, \text{\ and\ } \sigma_{z} = \frac{\sigma_{x}+i\sigma_{y}}{2}.
 \end{equation}
Whitfield et al.\cite{Lanyon2010towardsH2,Whitfield2011simulation}  and Seeley et al.\cite{Seeley2012}  thoroughly studied this mapping and used the Hamiltonian for the hydrogen molecule as an example system:
Using a minimal number of basis, only four spin orbitals indexed from $0$ to $3$ are involved in the above sum. They found the values of the one- and two-electron integrals by using a restricted Hartree-Fock calculation at an internuclear separation of 
$7.414 \times10^{-11} m$. Because of the overlap in the integral values, 
the Hamiltonian is reduced to the following form: 
\begin{equation}
\label{hamiltonianH2}
\begin{split}
\mathcal{H} =\ & h_{00}a_{0}^\dagger a_0 + h_{11}a_{1}^\dagger a_1+h_{22}a_{2}^\dagger a_2+h_{33}a_{3}^\dagger a_3,
\\ 
& +  h_{0110}a_{0}^\dagger a_{1}^\dagger a_{1} a_{0} + 
h_{2332}a_{2}^\dagger a_{3}^\dagger a_{3} a_{2} +
 h_{0330}a_{0}^\dagger a_{3}^\dagger a_{3} a_{0} 
\\
& + h_{1221}a_{1}^\dagger a_{2}^\dagger a_{2} a_{1}  +
(h_{0220}-h_{0202})a_{0}^\dagger a_{2}^\dagger a_{2} a_{0}
\\ & + 
(h_{1331}-h_{1313})a_{1}^\dagger a_{3}^\dagger a_{3} a_{1}
\\ &
+ (h_{0132})(a_{0}^\dagger a_{1}^\dagger a_{3} a_{2} +
a_{2}^\dagger a_{3}^\dagger a_{1} a_{0} )
\\ & +(h_{0312})(a_{0}^\dagger a_{3}^\dagger a_{1} a_{2} +
a_{2}^\dagger a_{1}^\dagger a_{3} a_{0} ).
\end{split}
\end{equation}
Using the mappings in Eq.\eqref{annihilationcreation}, 
the  Hamiltonian is  rewritten into the terms of Pauli matrices with the values of the coefficients given in Table \ref{table1} (taken from Eq.(80) of Ref.\cite{Seeley2012}):
\begin{equation}
\label{EqH2finalH}
\begin{split}
\mathcal{H} =& 
\beta_1I+
\beta_2\sigma_z^0 +
\beta_3\sigma_z^1 +
\beta_4\sigma_z^2 +
\beta_5\sigma_z^3 +
\beta_6\sigma_z^1 \sigma_z^0+
\beta_{7}\sigma_z^2\sigma_z^0  \\ &+ 
\beta_{8}\sigma_z^2\sigma_z^1 + 
\beta_{9}\sigma_z^3\sigma_z^0 +
\beta_{10}\sigma_z^3\sigma_z^1 +
\beta_{11}\sigma_z^3\sigma_z^2 \\ & +
\beta_{12}\sigma_x^3\sigma_x^2\sigma_y^1\sigma_y^0 +
\beta_{13}\sigma_x^3\sigma_y^2\sigma_y^1\sigma_x^0 +
\beta_{14}\sigma_y^3\sigma_x^2\sigma_x^1\sigma_y^0  \\ & +
\beta_{15}\sigma_y^3\sigma_y^2\sigma_x^1\sigma_x^0
\end{split}
\end{equation}
The simulation of this Hamiltonian within the phase estimation algorithm is done after generating a circuit equvailent of $e^{i\mathcal{H}t}$ through the Trotter-Suzuki approximation. 
As mentioned before, the complexity of the generated circuit is determined by the order of the approximation which affects the accuracy of the obtained ground state energy of the Hamiltonian from the phase estimation algorithm.

\begin{center}
\begin{table}
\caption{The values of the coefficients in Eq.\eqref{EqH2finalH}.\label{table1}}
%		        1 2 3 4
\begin{tabular}{|c|c|c|c|}
\hline
Coefficient & Value & Coefficient & Value\\ \hline
$\beta_1$  &	-0.8126	 &$\beta_{9}$ 	&0.1659	 \\
$\beta_2$ &	0.1712	  &$\beta_{10}$ 	&0.1205	 \\
$\beta_3$ &	0.1712	  &$\beta_{11}$ 	&0.1743	\\
$\beta_4$ &	-0.2228	  &$\beta_{12}$ 	&-0.0453	\\
$\beta_5$ &	-0.2228	 & $\beta_{13}$ 	&0.0453	\\
$\beta_6$ &	0.1686	&  $\beta_{14}$ 	&0.0453	\\
$\beta_{7}$& 	0.1205	 & $\beta_{15}$& 	-0.0453	\\
$\beta_{8}$ &	0.1659	&  $\kappa$ 	&20.117	\\
\hline
\end{tabular}
\end{table}
\end{center}

In our case, we first convert the Hamiltonian into the form of Eq.\eqref{EqBaseForm} by using $\kappa = 10\times||\mathcal{H}||_1$, which guaranties that $|\lambda_j/{\kappa}| \leq 0.1 \approx sin(0.1)$. 
 This results in total 16 terms for  $select(\bf{V})$, where $\kappa$ is taken as the coefficient for the additional identity and 
 the negative signs and $i$ are shifted to the terms inside $select(\bf{V})$ so that 
 all $\beta_j$s become positive real numbers (This is because $\sqrt{\beta_j}$s are involved in the construction of $\bf{B}$).   
 
 In the circuit, an ancilla of 4  qubits is required to control each term separately  to construct $select(\bf{V})$.
 As also mention in the previous section, in Ref.\cite{Mottonen}, a circuit  decomposition technique is described for circuits (called gray-coded networks) where each quantum gate on a qubit is controlled by $2^l$  distinct states of some other $l$ qubits.
It is shown that, the network can be decomposed into $2^l$ number of $CNOT$ and $2^l$ number of single gates by using some form of Hadamard transformation.
The circuit for $select(\bf{V})$ consists of 4 such networks (one network for per qubit in the system). Each network is controlled by $logL$ qubits in the ancilla.
Since each decomposed network will require $L$ number of $CNOT$s and $L$ number of single quantum gates, the circuit complexity of $select(\bf{V})$ is bounded by $O(nL)$, where $n$ is the number of qubits in the system. 
As a result, the for circuit $H_2$ will require 64 $CNOT$ gates.
In the previous section, the circuit complexity of $\bf{B}$ is given by $O(L)$, which makes the whole circuit complexity $\approx{100}$ gates.
In the phase estimation, since we have also a phase qubit added to the control, the required number of gates for $select(\bf{V})$ will be $\approx128$ gates. 

\begin{figure*}[h]
\includegraphics[width=7in]{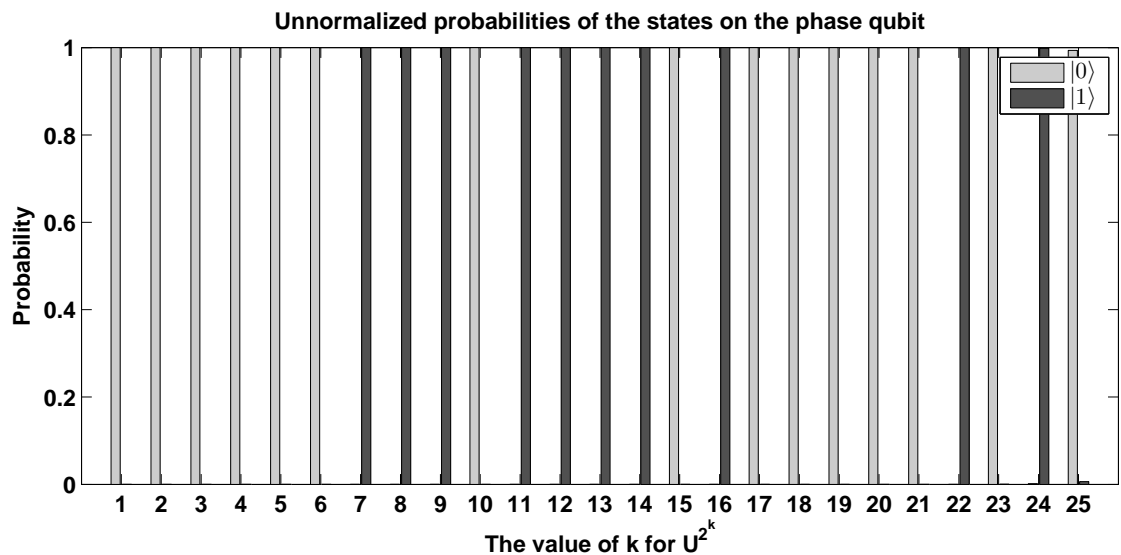}
\caption{The probabilities on the phase qubit with the amplitude amplification applied only to $\bf{U}^{(1)}$: i.e. $\bf{Q}^6\bf{U}^{(1)}$. The power of the matrix is estimated by direct application: for square it is $\bf{Q}^6\bf{U}^{(1)}\bf{Q}^6\bf{U}^{(1)}$. Also note that the probabilities are not normalized after the application of $\bf{P}$. 
\label{FigwithAA1}
}
\end{figure*}

\begin{figure}[h]
\includegraphics[width=3.5in]{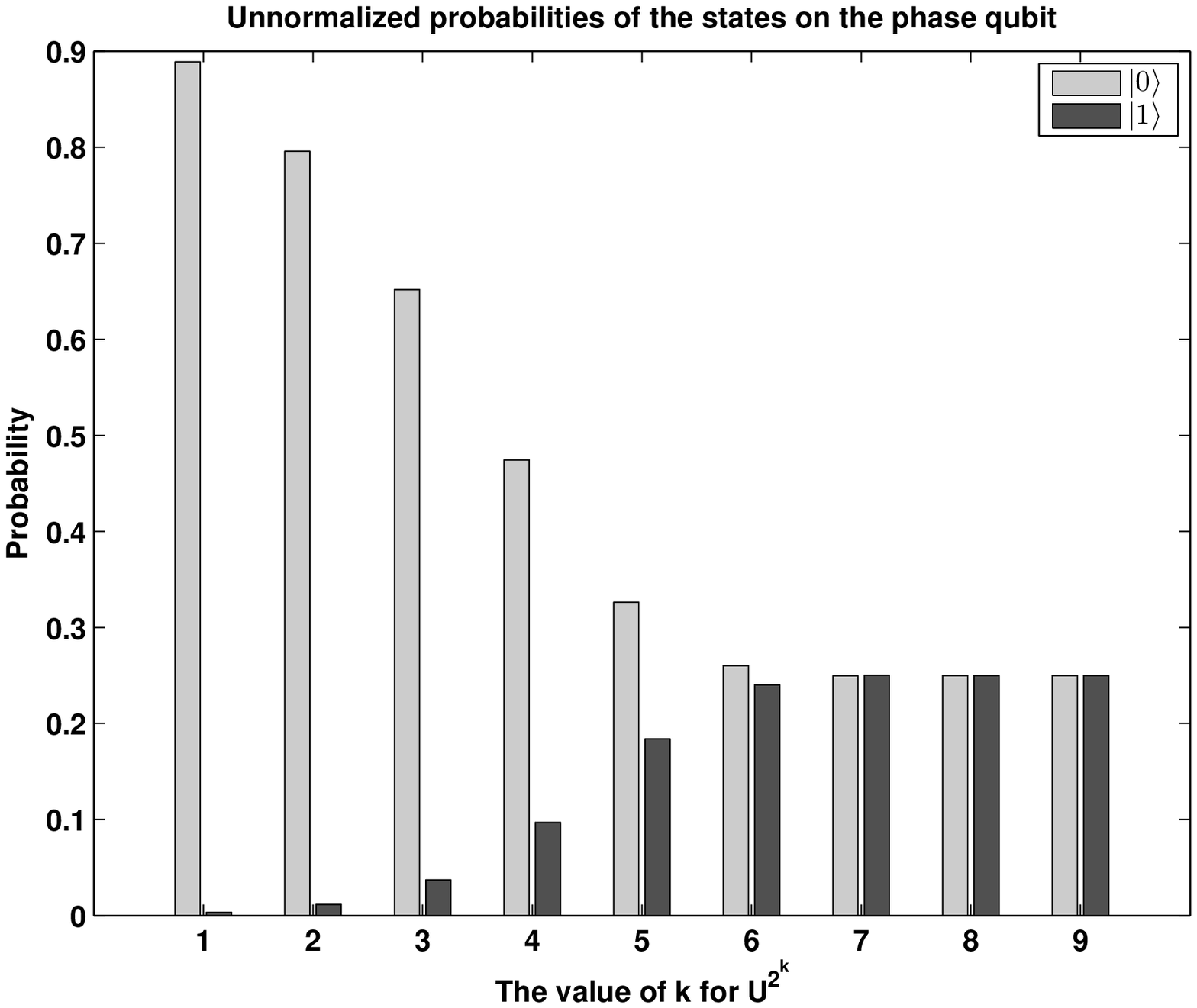}
\caption{The probabilities on the phase qubit without the amplitude amplification applied to $\bf{U}^{(1)}$. Note that the probabilities are not normalized after the application of $\bf{P}$. \label{FigwithoutAA}}
\end{figure}

\begin{figure}[h]
\includegraphics[width=3.5in]{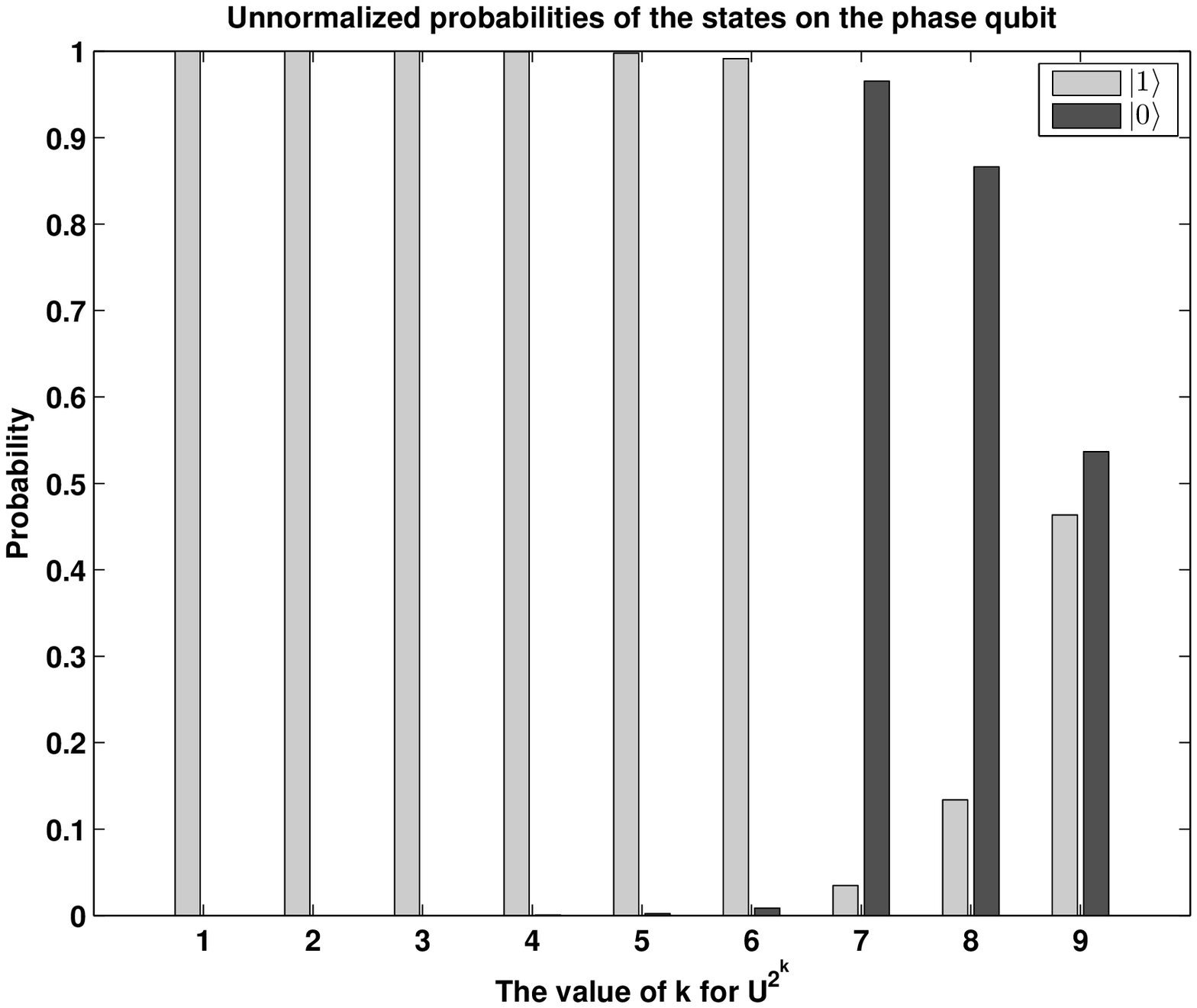}
\caption{The probabilities on the phase qubit with the amplitude amplification applied only to $\bf{U}^{(1)}$. Note that the probabilities are not normalized after the application of $\bf{P}$. 
\label{FigwithAA}
}
\end{figure}

The simulation is done in three different ways:
\begin{itemize}
\item In the first case, the powers are estimated through the successive application of $\bf{\widetilde{U}^{(1)}}$ given in Eq.\eqref{EqU(1)AA}. Here the amplitude amplification operator $\bf{Q}$ given in Eq.\eqref{EqOAAmp} is applied six times to obtain a maximum probability. The phase estimation algorithm is run for 25 iterations and the measured unnormalized probabilities for the phase qubit are drawn in Fig.\ref{FigwithAA1}.
The obtained phase from the bit values is 0.014603. From $\arcsin(\text{Im}(e^{-i2\pi 0.014603}))\times\kappa$, the value of the ground state energy is found as -1.845843. 
Since the system size is small, the ground state energy of the Hamiltonian has been also   classically computed as $-1.851046$ via eigen-decomposition. 
This shows an error value $0.0052$ in our estimation caused by the error in the estimation of the powers of the matrices through Eq.\eqref{EqU(1)AA} and the error in the approximation $sin(\lambda_j/\kappa)\approx \lambda_j/\kappa$.

\item In the other cases, the power of the matrices is computed through Eq.\eqref{EqPowerofH} 
by adding one more qubit to the ancilla in each iteration.
Here, because of the computational difficulty-the system size for 9 iteration  requires 17 qubits, we  present the output of the phase qubit for only 9 iterations of the phase estimation without and with the amplitude amplification applied to $\bf{U^{(1)}}$ respectively 
in Fig.\ref{FigwithoutAA} and Fig.\ref{FigwithAA}.  
In comparison to classically computed value, the simulation results produce the correct eigenvalue  if the previous bit values are provided; otherwise, it generates an estimate of the eigenvalue, as expected, with an error bounded by $O(\kappa/ 2^{-9})$. 
As seen in Fig.\ref{FigwithoutAA}, without the amplitude amplification, depending on the value of $k$
 in $\bf{U}^{2^k}$,
 the probability difference between 1 and 0 on the phase qubit diminishes and the total success probability approaches to 0.5 (the probability of the part of the ancilla system simulating $\wmc{H}$).
 As shown in Fig.\ref{FigwithoutAA} with the same $\kappa$ value,  the total probability can be maximized by the application of the amplitude amplification only to $\bf{U^{(1)}}$.
  The probability difference now survives for 9 iterations. 
  This can be further improved by using oblivious amplitude amplification algorithm in the subsequent iterations when the probability difference starts to drop and 
  changing the value of $\kappa$. 
  Note that a larger $\kappa$ will  require more iterations to obtain a higher precision.
 Also note that bit values and the probabilities are affected by the number of iterations because the iterative phase estimation algorithm uses a rotation gate whose angle determined by the bit values measured in the previous iterations.
\end{itemize}

\section{Conclusion}
In this paper, we have described a framework to estimate the eigenvalues of Hamiltonians in the phase estimation algorithm without using the time evolution operator, the exponential, of the Hamiltonian. 
We have shown how to find the powers of the matrices necessary in the phase estimation algorithm by successive applications.
We have analyzed the circuit implementation of the whole framework in terms of classical and quantum complexities and showed that
the framework provides an efficient way to estimate the eigenvalues of Hamiltonians which are written in terms of sums of simple unitary matrices (here, ``simple" means a unitary matrix which can be implemented through a few number of one- and two-qubit quantum gates.).
In addition,  we have used the Hamiltonian of hydrogen molecule as an example system and showed how to estimate its ground state energy through the described method. 
We believe this framework can be used efficiently for many eigenvalue related problems such as the finding ground state energy of quantum systems.

\pagebreak
\bibliography{ref}
\end{document}